\documentclass[aps,prb,twocolumn,amsmath,amssymb]{revtex4}
\usepackage{graphics}
\usepackage{amstext,amsopn,amsmath,amssymb}
\usepackage{epsfig}

\begin{document}

\title{Soft particle model for block copolymers}

\author{F. Eurich$^1$, A. Karatchentsev$^1$, J. Baschnagel$^2$, W. Dieterich$^1$, P. Maass$^3$}

\affiliation{$^1$Fachbereich Physik, Universit\"at Konstanz,
D--78457 Konstanz, Germany} \affiliation{$^2$ Institut Charles
Sadron, 6 rue Boussingault, Strasbourg, Cedex 67083, France}
\affiliation{$^3$ Fakult\"{a}t f\"{u}r Mathematik und Naturwissenschaften,
Technische Universit\"{a}t Ilmenau, Postfach 100565, 98648 Ilmenau,
Germany }


\begin{abstract}

A soft particle model for diblock (AB) copolymer melts is
proposed. Each molecule is mapped onto two soft spheres built by
Gaussian $A$- and $B$-monomer distributions. An approximate
analytical expression for the joint distribution function for the
distance between both spheres and their radii of gyration  is
derived which determines the entropic contribution to the
intramolecular free energy. Adding a mean-field expression for the
intermolecular interactions, we obtain the total free energy of
the system. Based on this free energy,  Monte Carlo simulations
are carried out to study the kinetics of microphase ordering in
the bulk and its effect on molecular diffusion. This is followed
by an analysis of thin films with emphasis on pattern transfer
from walls with a periodic structure. It is shown that the level
of coarse graining in the soft particle model is suitable to
describe  structural and kinetic properties of copolymers on
mesoscopic scales.

\end{abstract}
\maketitle

\section*{1. Introduction}

Coarse graining by elimination of irrelevant degrees of freedom is
a central problem in many areas of statistical physics, especially
when modelling diffusion and the phase kinetics in polymer
systems.\cite{Bi94} An important class of processes is phase
separation in polymer melts with incompatible components, which
occur on time and length scales many orders of magnitude larger
than those related to the motion of individual monomers. A
detailed understanding of such processes is required in
present-day attempts to utilize polymers in the design of new
materials and in the production of tailored micro- or nanoscale
structures by
 self-organization.\cite{Bi94,Boe98}

From the theoretical viewpoint this poses the problem to eliminate
internal degrees of freedom of polymer chains, and to seek for a
coarse grained description of the kinetics in terms of a small
number of collective degrees of freedom. In order to cover effects
of chain length, monomer interactions, heterogeneities in monomer
sequences e.g. in the case of copolymers, and interactions with
walls in confined systems,\cite{Ba95} such coarse grained theories
should be based on appropriate microscopic input.

A promising approach is to represent polymer chains in their coil
 conformation in dense systems simply by interpenetrating ``soft
 particles''. This soft-particle picture should be applicable for
 length scales comparable to or larger than the radius of gyration
 $R_G$ and for time scales of the order of the diffusion time $\tau_D
 \simeq R_G^2/D$, where $D$ denotes the center of mass diffusion
 constant. The soft geometrical object assigned to a polymer can be
 quite general and should be characterized by only a few parameters.
 In the simplest case, polymers are described by soft spheres with
 effective interactions that can be derived by an explicit elimination
 procedure for monomer degrees of freedom.\cite{Likos} Slightly more
 complex geometrical objects are soft ellipsoids, which allow one to
 take into account orientational effects and anisotropic shape
 deformations. For homopolymers the first soft-ellipsoid model was
 developed in Ref. \cite{Mu98} based on microscopic input derived from
 a bead spring model. This model includes the excluded volume effect
 into the effective intramolecular free energy. As a consequence,
 tuning of interaction parameters with control parameters
 (temperature, degree of polymerization) is necessary to capture
 correctly the screening effect responsible for the quasi-ideal chain
 behavior in dense systems. It was shown that this adaption of
 effective interaction parameters can be avoided by using the Gaussian
 chain model as basis for the soft ellipsoid model,\cite{Eu01} at the
 expense that the excluded volume effect in this model has to be
 included in a mean field type manner in the self-interacting term of
 the free energy. Based on the Gaussian ellipsoid model, structural
 and kinetic properties of both one-component melts and binary
 mixtures were studied successfully for bulk systems. In a subsequent
 extension to confined systems it was shown how spinodal decomposition
 in polymer blends becomes modified in thin films, including
 situations of periodically patterned walls.\cite{Eu02}

Soft particle models become increasingly valuable in the
description of complex matter systems, and also analytical
techniques can be used to relate the input parameters to
microscopic properties.\cite{YabenkoA,YabenkoB}

In this paper we propose a soft particle model for block
copolymers. Phase separation in these systems  proceeds under the
constriction of a chemical link between incompatible polymer
blocks.\cite{Ba99,Ma02} Depending on the relative amounts of
different blocks, a variety of ordered microphases can emerge on
length scales that are tunable by the respective degrees of
polymerization. Because of these unique features block copolymers
offer a promising tool for the fabrication of nanoscale devices
as, for example, by the nanolithography
technique.\cite{Srinivas,Zschech}

Following the early fundamental work on the derivation of a Landau
theory by Leibler\cite{Lei80} and of fluctuation corrections by
Ohta and Kawasaki\cite{Oh86} and subsequently by Fredrickson and
Helfand,\cite{Fre87} a progressive understanding of the phase
diagram of diblock-copolymers with variable length ratio of the
two blocks has been developed. Important features like the static
structure factor and the scaling of the microphase periodicity
with chain length were examined  by the lattice Monte Carlo
technique.\cite{Frie91,Hof1,Hof2} Further simulation studies
provided insight into the detailed changes of chain conformation
near phase boundaries.\cite{Gre96,MuGre98,Mu99}  the internal
energy and entropy,\cite{ASchulz02} as well as transport
properties, e.g. center-of-mass diffusion of chains and their
modification by the order-disorder transition.\cite{Hof2,Mu99}
More sophisticated methods are the ``dissipative particle
dynamics'' (DPD) that includes hydrodynamic forces between
effective beads of polymers,\cite{Gro98,Gro99} and methods
treating the entangled state of copolymers.\cite{Masubuchi}

On the other hand, for studying the far-from-equilibrium dynamics
and large scale structures, evolution equations for monomer
densities have been developed. One approach is to formulate
time-dependent Ginzburg-Landau (TDGL)
equations.\cite{Mau97,Nono03,Chakrabarty,Fra97} A powerful
extension thereof is the selfconsistent (SFC) field
method,\cite{Matsen:1995,Krausch} which proved successful in
predicting large scale phase morphologies.

Complementary to these developments is the soft particle model
proposed here. Motivated by the results of Refs. \cite{Eu01,Eu02}
we explore the possibility to map the internal molecular degrees
of freedom onto only a few parameters, allowing straightforward
physical interpretation. In other words, we attempt to
coarse-grain as far as possible while keeping the most important
structural characteristics of individual molecules.  Because of
its simplicity, this model should enable one to treat the time
evolution of ordering structures, including surface-induced
structures, in an efficient manner. Unlike TDGL or SCF theories,
it retains orientational effects and shape fluctuations of
individual molecules,  features which again will be important when
treating surface effects. Moreover, a model of this type could
form the basis for treating the phase behavior of more complex
molecules.

Our  paper is organized as follows. First, we develop in section 2
a  soft-particle model of (AB)-diblock copolymers, where each
block is represented by a Gaussian sphere. Their radii of gyration
$R_X \, (X = A,B)$ and the vector $\vec r$ connecting their
midpoints are the only parameters related to the internal degrees
of freedom of a molecule. Test calculations for the structure
factor, the ordering transition, the scaling of the lamellar
distance, stretching of chains, as well as diffusion properties in
the bulk are presented in section 3. Thereby we demonstrate that
this Gaussian disphere model (GDM) reproduces known bulk
properties in a satisfactory manner. This motivates us to  treat
in section 4 equilibrium and kinetic properties in thin film
geometries. Major questions concern the wall-induced microphase
separation both for neutral walls and walls that attract  one
component, and the dynamics of pattern transfer from  prepatterned
walls into the film.  Finally, some concluding remarks and
directions for further research are presented in section 5.

\section*{2. Gaussian disphere model for block copolymers}

\begin{figure*}[ht]
\centering\includegraphics[width=0.60\textwidth]{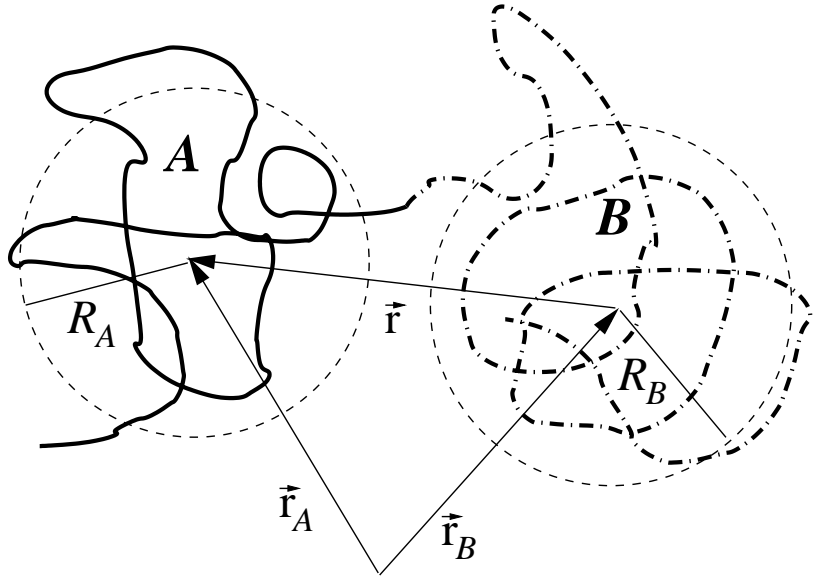}
\caption{Schematic illustration of the Gaussian disphere model.}
\label{F1}
\end{figure*}

Following the basic idea in previous work \cite{Eudiss} we propose
here to represent each block in an (AB)-diblock copolymer molecule
by one soft sphere with radius of gyration $R^X \, (X = A$ or $
B$), as illustrated in Fig.´~\ref{F1}. The molecules' orientation
is given by the vector $\vec r = \vec r^A - \vec r^B$ connecting
the two centers of the spheres. Its magnitude $r = |\vec r|$
determines the stretching of the molecule under AB-repulsion. The
three quantities $R^A, R^B$ and $\vec r$ are the parameters that
represent the internal degrees of freedom of one molecule. In
order to express configurational fluctuations of a molecule in
terms of these parameters, we have to specify
\begin{itemize}
\item[{i)}] the joint probability densities $P(\vec r, R^A,\, R^B)$
for given number of bonds in the $A$ and $B$ block, $N_A$ and $N_B$,
respectively.
\item[{ii)}] the conditional monomer densities $\rho_X(\vec x|\vec
r,R^A,R^B)$ of the block $X$ written in terms of the coordinate
$\vec x $ relative to the center $\vec r^X$ of the sphere. The
normalization condition is
\begin{equation}\label{rhox}
\int \rho_X(\vec x|\vec r,R^A,R^B) {\rm d}^3 x = N_X +1 \,.
\end{equation}
\end{itemize}
These densities are to be derived from a
microscopic model of an isolated chain.  For the free energy of an
ensemble of $M$ molecules with fixed values $N_A$ and $N_B$ we write
\begin{equation}\label{F-in-in}
F = F_{\rm intra} + F_{\rm inter}\,,
\end{equation}
where
\begin{equation}\label{Fintra}
  F_{\rm intra} = - k_B T \sum_{i = 1}^M \ln P(\vec r_i,R_i^A,R_i^B)
\end{equation}
is the intramolecular part in the absence of intermolecular
interactions. For the latter, contact interactions of strength
$\epsilon_{XY}$ among the different blocks $(X,Y = A$ or $B$) are
considered here,
\begin{equation}\label{F-inter}
  F_{\rm inter} = \frac{1}{2}\, \sum^M_{i,j=1}\, \sum_{X=A,B}\,\sum_{Y=A,B}
  \epsilon _{XY}\, b^3 \int d^3y\, \tilde\rho^X_i(\vec y)\,
  \tilde\rho^Y_j(\vec y)\, ,
\end{equation}
where $b^3$ is a contact volume. The parameter $b$ sets the
microscopic length scale of the model and is used as length unit,
$b=1$. The density $\tilde\rho_i^X$ denotes the $X$-monomer
density in the laboratory frame,
\begin{equation}\label{rho-ix}
  \tilde\rho^X_i(\vec y) = \rho_X(\vec y-\vec r^{\,X}_i|\vec r_i, R^A_i,
  R^B_i)\, .
\end{equation}
 Note that we have included the self-interaction terms
$i=j$ in eq.~(\ref{F-inter}), which implies that we should consider
entropic contributions only in $F_{\rm intra}$.

To complete the definition of the free energy, we must specify the
input functions $P$ and $\rho_X$. Following Ref.~\cite{Eu01}, we
propose to use Gaussian chains as microscopic input, with a few
additional approximations that lead to simple analytic
expressions. First we consider $P$. Here our assumption is that
$R^A$ and $R^B$ are uncorrelated, with distributions $P_X(R^X)$
obtained from homogeneous Gaussian chains. Thus we can write
\begin{equation}\label{P-rAB}
  P(\vec r,R^A,R^B)=P_A(R^A)\,P_B(R^B)\,W(\vec r \, | R^A,R^B)
\end{equation}
with
\begin{equation}\label{Px-rx}
P_X(R^X)\sim
\frac{1}{\sqrt{N_X}}\,p\left(\frac{R^X}{\sqrt{N_X}}\right)\,, \,
\int^\infty_0 du\, p(u)=1
\end{equation}
for large $N_X$. An accurate representation of the scaling function
is \cite{Eu01}
\begin{equation}\label{tilde-P}
  p(u) = \frac{1}{u K_0 (2d)}\exp \left(- \frac{u^2}{a} -
  d^2 \frac{a}{u^2}\right)\,,
\end{equation}
where $K_0(z)$ denotes the modified Bessel function of order zero
and the coefficients $a$ and $d$ are known from fits to the exact
moments of second and fourth order in the limit of large $N_X $
$(\langle (R^X)^2 \rangle \simeq N_X/6$; $\langle (R^X)^4 \rangle
\simeq (19/540)N_X^2$). For the conditional probability in
(\ref{P-rAB}), we make a Gaussian ansatz with respect to $\vec r$,
\begin{equation}\label{P-RA-RB}
  W(\vec r\, |R^A,R^B) = \left(\frac{3}{2\pi\langle \vec
  r^{\,\,2}\rangle_R}\right)^{3/2}
  \exp\left(-\frac{3\vec r^{\,\,2}}{2\langle\vec r^{\,\,2}\rangle_R}\right)
\end{equation}
where $\langle\vec r^{\,\,2}\rangle_R=2[(R^A)^2 + (R^B)^2]$ is the
variance of $\vec r$ for given $R^A$ and $R^B$.

\begin{figure*}[ht]
\centering\includegraphics[width=0.70\textwidth]{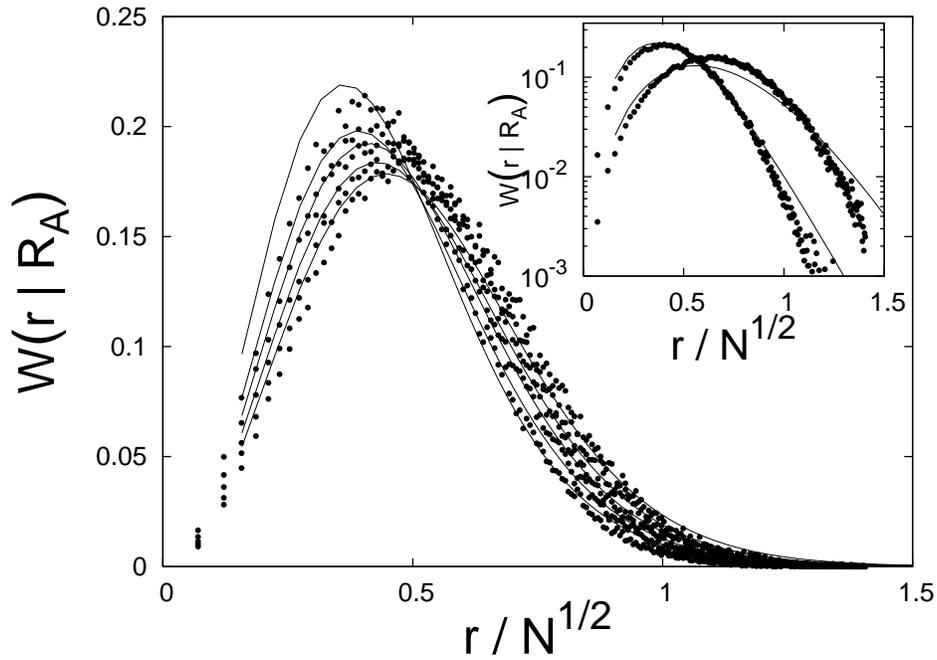}
\caption{Comparison between the conditional probability $W(r|R^A)$
based on the Gaussian approximation (\ref{P-RA-RB}) (continuous
curves) with Monte Carlo data for symmetric chains with $N = 100$.
To display the dependence on $R_A$, averages have been taken over
eight successive $(R^A)$-intervals, see text. Different curves
refer to the first five of these intervals.  With increasing
$R^A$-values, distributions $W(r|R^A)$ shift towards larger
$r$-values. The inset shows, in a semi-logarithmic representation,
the  results for the  smallest and largest of these eight
$(R^A)$-intervals,  confirming that the Gaussian approximation is
satisfactory as long as $r \lesssim N^{1/2}$. } \label{F2neu}
\end{figure*}

As a test of (\ref{P-RA-RB}) we compare with Monte Carlo
simulations the conditional probability $W (r|R^A)$ obtained from
(\ref{P-RA-RB}) by integrating over the orientations of $\vec r$
with  $r = |\vec r|$ fixed, and by averaging over $R^B$. Regarding
its dependence on $R^A$, we took averages over each of eight
successive intervals on the $R^A$ axis chosen such that they have
equal weights 1/8 with respect to the distribution (\ref{Px-rx}).
Results are shown in Fig.~\ref{F2neu} by the continuous curves,
together with the simulation data. With increasing $R^A$, peak
positions move to the right,  showing that chain realizations with
large $R^A$ imply larger stretchings $r$. The semilogarithmic plot
in the inset confirms satisfactory agreement between the data and
the Gaussian approximation.

Second, we turn to the conditional monomer density $\rho_X$. As
mentioned already, we assume a spherical shape and ignore any
dependence on $\vec r$ and on the size of the opposite block, i.e.
$\rho_X = \rho_X (\vec x |R^X)$.  Then for large $N_X$ and for
typical $R^X$ (excluding highly stretched chains) one can show
that the  scaling form
\begin{equation}\label{rho-X}
  \rho_X (\vec x |R^X)\sim \frac{N_X + 1}{(R^X)^3}f\left(\frac{x}{R^X}\right)
\end{equation}
 with $f(v) =(3/2\pi)^{3/2}\exp (-3v^2/2)$ well agrees with Monte Carlo simulations for isolated chains.
 Using (\ref{rho-X}) as approximation in the GDM directly follows the analogous
 reasoning in the soft ellipsoid model, described  in detail in Ref.~\cite{Eu01}.
 To complete the description of the GDM, we give the expression for the radius of gyration $R_G$ of the total chain,
 to be derived by simple geometric considerations
\begin{equation}\label{radius}
R_G^2 = f_A [R^A]^2 + f_B [R^B]^2 + f_A f_B \vec r^{\,2}\, ,
\end{equation}
 where $f_X = N_X/N$, and $N = N_A + N_B$. It is clear that this type of model can be extended to
more complicated structures like triblock copolymers, but we will not
pursue this here.

So far, for given external parameters $N_X$, $M$ and $V$, we have
constructed a free energy $F$ that depends on the variables
$R^X_i$ and $\vec r^{\,X}_i$ (giving $\vec r_i=\vec r^{\,A}_i-\vec
r^{\,B}_i$). To model the kinetics, these varaiables are changed
in a Monte Carlo process. Two types of elementary moves are
considered: {\it (i)} block translations $\vec r^X_i \rightarrow
\vec r^X_i+\Delta\vec r^X_i$ where the components of $\Delta\vec
r^X_i$ are drawn from a uniform distribution in the interval
$[-\Delta r_{\rm max}^X/2,\Delta r_{\rm max}^X/2 ]$ with $\Delta
r^X _{\rm max} = (3/4)(N_X/6)^{1/2}$, and {\it (ii)} size changes
$R^X_i \rightarrow R^X_i+ \Delta R^X_i$, where $\Delta R^X_i$ is
uniformly distributed in the interval $[-\Delta R_{\rm
max}^X/2,\Delta R_{\rm max}^X/2]$ with $\Delta R^X _{\rm
max}=0.5(N_X/6)^{1/2}$ under the additional constraint $R^X_i>0$.
All elementary steps have the same a priori probability.
Acceptance probabilities are chosen according to the Metropolis
algorithm based on the free energy (\ref{F-in-in}).

Results are presented in terms of the Flory-Huggins type parameter
$\chi=(2\epsilon_{AB}-\epsilon_{AA}-\epsilon_{BB})/k_{\rm B}T$ and
we will use $k_{\rm B}T$ as energy unit in the following, $k_{\rm
B}T=1$. All calculations are performed for
$\epsilon_{AA}=\epsilon_{BB}  = 1$ and a total mean density of
monomers $\overline\rho=0.85$.  Unless otherwise said, symmetric
mixtures with a fraction $f_{\rm A}=0.5$ of A monomers are
considered and a chain length $N=120$ is chosen.

\section*{3. Bulk properties}

In this section we test the GDM against bulk properties of
chain-like copolymeres, a subject well known from extensive model
studies.\cite{Lei80,Oh86,Fre87,Frie91,Hof1,Hof2,Gre96,MuGre98,Mu99,ASchulz02,Gro98,Gro99}
We show that despite its simplicity the GDM accounts for a
remarkable set of both structural and diffusion properties
connected with microphase separation.  In all bulk simulations
periodic boundary conditions are used and the volume $V$ of the
cubic simulation box is chosen to contain in total $M=4000$
molecules. Averages are typically performed over 10 independent
simulation runs.

\subsection*{3.1 Structure}

\begin{figure*}[ht]
\centering\includegraphics[width=0.60\textwidth]{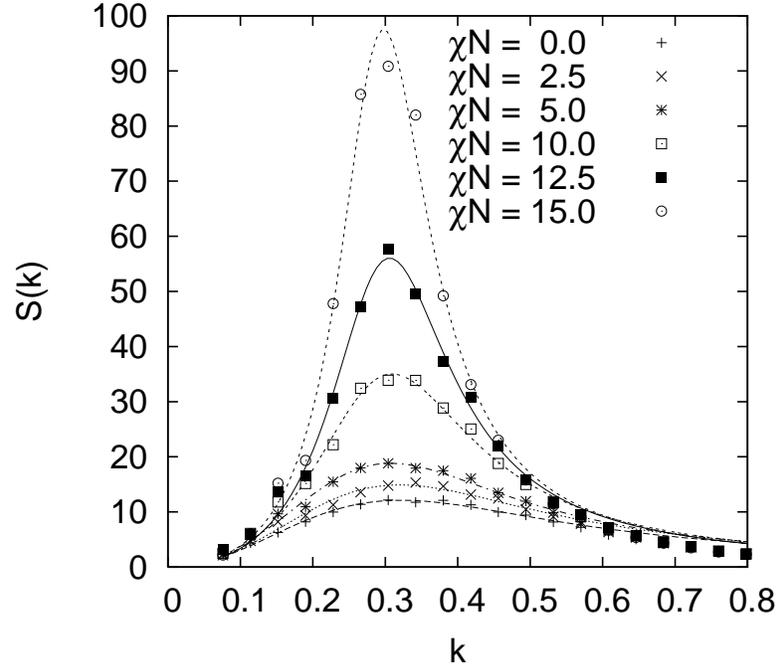}
\caption{Simulated structure factor $S( k )$  in the disordered
phase for different $\chi N$. Continuous curves are fits to Eq.
(\ref{NS1}).}
 \label{F2}
\end{figure*}

A direct measure of short- and long-range ordering is the static
structure factor $S(\vec k )$, which we define through the density
correlation function of A-monomers,
\begin{equation}\label{svecq}
S(\vec k ) = \frac{V}{MN_A}\int d^3 x \, \langle \rho_A (\vec x
\,) \rho_A (0)\rangle e^{i \vec k\cdot\vec x}\,.
\end{equation}
In Fig.~\ref{F2}, data points of the spherical average of
(\ref{svecq}) are displayed for symmetric chains in the disordered
phase. Continuous curves are fits to the expression
\cite{Lei80,Frie91}
\begin{equation}\label{NS1}
NS^{-1} ( k ) = \frac{1}{\alpha}[F(k \tilde R_G) - \delta]
\end{equation}
with
\begin{equation}\label{Fx}
F(x) = \frac{x^4}{2}\left( \frac{x^2}{4} + e^{-x^2/2} -
\frac{e^{-x^2}}{4} - \frac{3}{4}\right).
\end{equation}

Equation (\ref{NS1}) generalizes the form of the structure factor
from Leibler's random phase approximation, which in our notation
amounts to setting $\alpha = 1/2$, $\tilde R_G = R_G$ and $\delta
= 2 \chi N$. In this theory, lamellar ordering sets in at the
critical value $(\chi N)^{\rm L}_c = 10.5$, connected with a
divergence of $S(k)$ at $k =k_\star$ with $k_\star R_G=1.95$.
Following previous work \cite{Frie91,Mu99,Hof1} $\alpha, \tilde
R_G$ and $\delta$ are regarded as fit parameters, which we
determine from the behavior of $S( k )$ around its main peak at
$k_\star$. Their values are given in Table 1. In the vicinity of
the peak and towards small $k$ a good fit is achieved, whereas at
larger $k$ the data points fall below the continuous curve, see
Fig.~\ref{F2}. This is to be expected because the intramolecular
connectivity is not taken into account by assuming Gaussian
monomer densities within each block.

\begin{figure*}[ht]
\centering\includegraphics[width=0.60\textwidth]{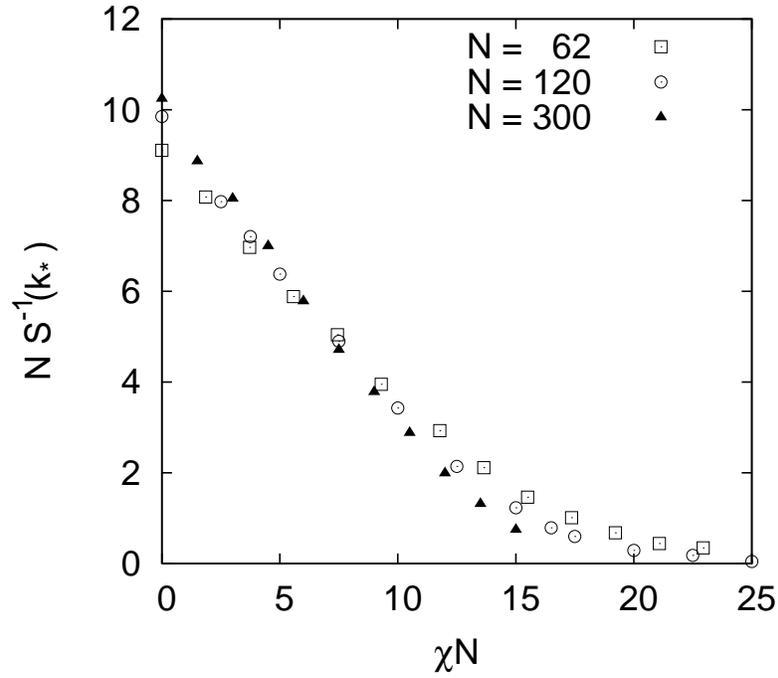}
 \caption{Normalized inverse maximum of the structure factor
 versus $\chi N$ for three different chain lengths.}
 \label{F3}
\end{figure*}

The physical meaning of the parameters $\alpha$, $\tilde R_G$, and
$\delta$ is to allow for a shift in the peak position $k_\star$
relative to the Leibler value \cite{Lei80} and a deviation of the
maximum value  $S(k_\star)/N$ from scaling with $\chi N$. Indeed,
our results in Fig.~\ref{F2} reflect a downward shift in $k_\star$
with increasing $\chi$, starting with $k_{\star}R_G \simeq 1.8$
for $\chi = 0$. Qualitatively, this downward shift agrees with
previous calculations,\cite{Mu99,Hof1} but the effect is smaller
than in these works. Moreover, when plotting $N S^{-1}(k_{\star})$
from calculations with different $N$ against $\chi N$, see
Fig.~\ref{F3}, linear extrapolation to zero yields an instability
of the disordered phase at a critical value $(\chi N)_c$ that
increases with decreasing chain length $N$. This trend is
qualitatively consistent with the Fredrickson and Helfand
theory,\cite{Fre87} predicting $(\chi N)^{\rm FH}_c - (\chi
N)_c^{\rm L} \propto N^{-1/3}$. In our model the critical value
for the longest chains $(N = 300)$ is roughly estimated as $(\chi
N )_c \simeq 12$ to 14.

\begin{figure*}[ht]
 \centering\includegraphics[width=0.50\textwidth]{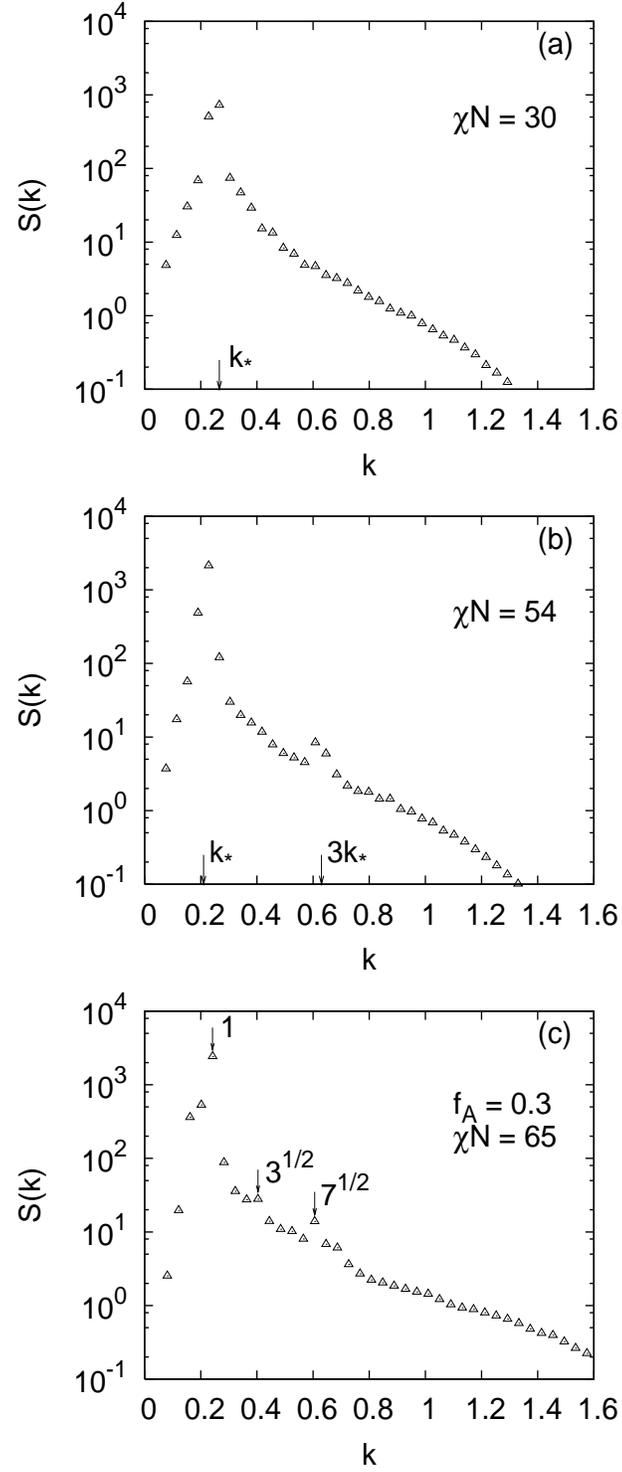}
\caption{Spherically averaged structure factor in ordered phases.
Lamellar phase  with (a) $\chi N = 30$ (weak segregation) and (b)
$\chi N = 54$ (strong segregation), displaying the third order
peak. (c) Cylindrical phase with $f_A = 0.3$, $\chi N = 65 \, (N =
100)$, with marked higher order peaks.}
 \label{F4}
\end{figure*}

\begin{figure*}[ht]
\centering\includegraphics[width=0.80\textwidth]{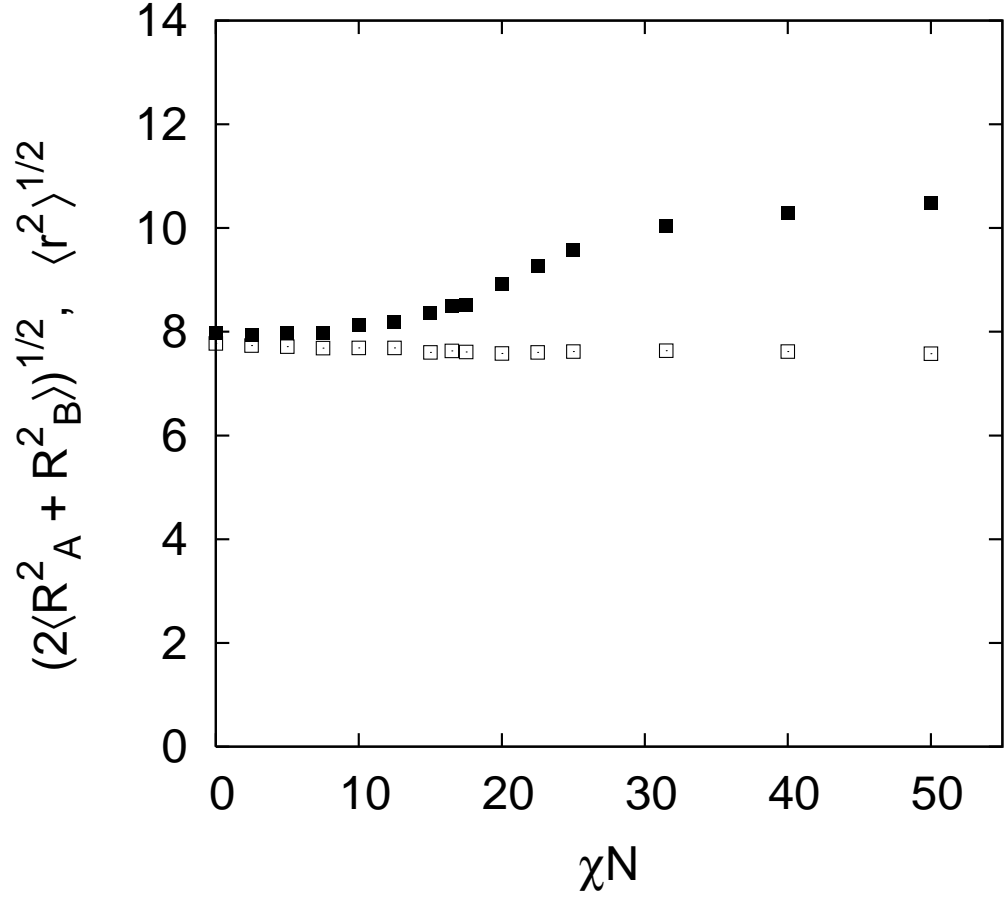}
 \caption{Averaged stretching parameter $ r $ (full symbols)
compared to  averaged radii of gyration $ \langle (R^A)^2
\rangle^{1/2} = \langle (R^B)^2 \rangle^{1/2}$ (light symbols),
cf. Eq. (\ref{P-RA-RB}),   versus $\chi N$.} \label{F5}
\end{figure*}

Ordered structures spontaneously forming under a quench from the
disordered state to a value $\chi N$ above but still close to $
(\chi N)_c$, display a multidomain pattern and weak segregation,
i.e.\ a smooth variation of the respective monomer densities when
passing from A-rich to B-rich domains. Figure \ref{F4}a
exemplifies the structure factor for $\chi N = 30$. The position
of the sharp peak at $k_\star = 2 \pi/\lambda $ reflects the
lamellar periodicity $\lambda$. Obviously, $k_\star$ is smaller
than in the disordered state of molecules with the same chain
length, due to their stretching under alignment. In the GDM, chain
stretching is described by the parameter $r = \langle \vec
r^{\,\,2}\rangle^{1/2}$. Its dependence on $\chi N$ (with $N$
fixed) is plotted in Fig.~\ref{F5}, showing a pronounced increase
in the vicinity of the ordering transition and a subsequent weaker
increase as $\chi N$ rises further. For $\chi = 0$, one recovers
$r^2= 2[\langle (R^A)^2 \rangle + \langle (R^B)^2\rangle ]$, see
section~2. The averaged radii of gyration $\langle
(R^X)^2\rangle^{1/2}$ of individual blocks are practically
unaffected by changing $\chi$.

Going to strong segregation, larger scale oriented lamellae with
sharp interfaces develop. This leads to higher-order peaks in
$S(k)$ seen in Fig.~\ref{F4}b, where we plotted the structure
factor for $\chi N = 54$. Note the absence of the second order
peak due to the arrangement of blocks inside the domains. We also
carried out bulk simulations for asymmetric chains. The existence
of a cylindrical phase is exemplified by Fig.~\ref{F4}c for $f_A =
0.3$, reflected by superstructure peaks as marked in the figure.

\begin{figure*}[ht]
\centering\includegraphics[width=0.80\textwidth]{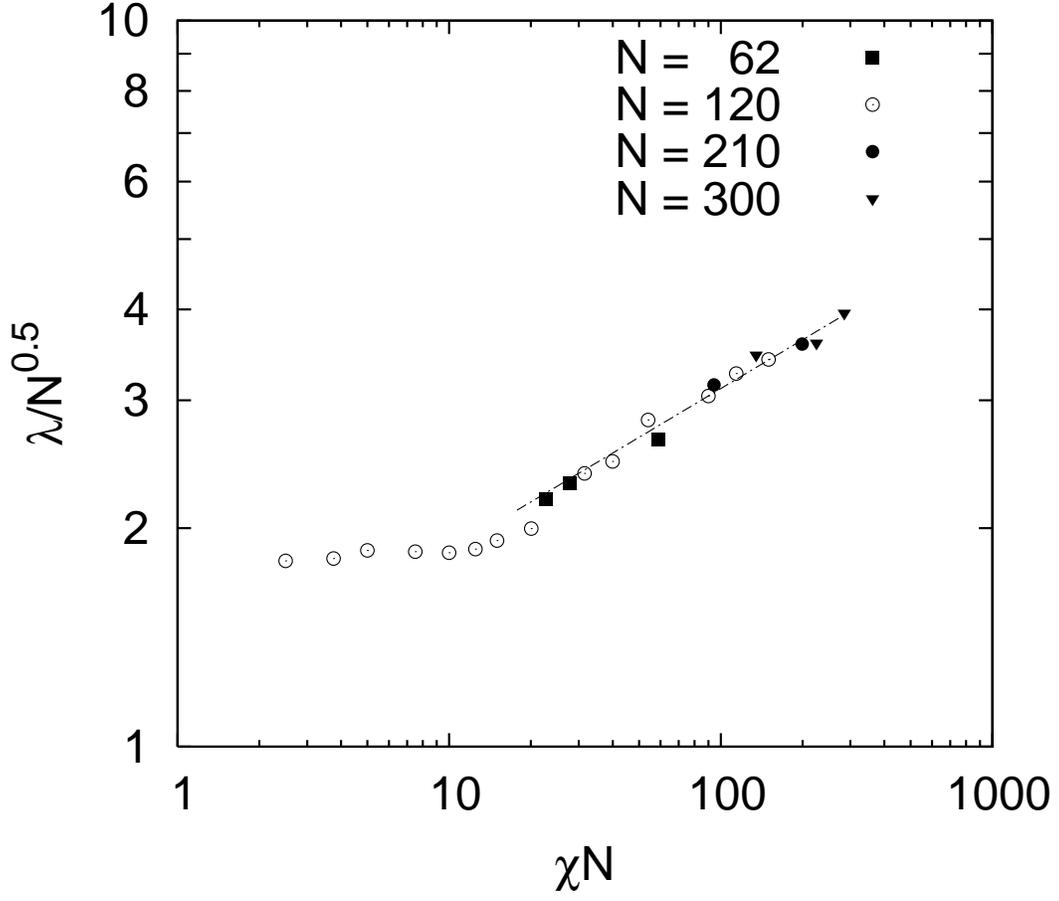}
 \caption{Scaling plot of lamellar distance $\lambda$
depending on $\chi$ and $N$ extending to  the strong
segregation-regime. The dashed straight line has a slope $n \simeq
0.22$. Data points for $N = 120$ are continued to the disordered
phase. In these simulations, averages were taken over three
independent runs.}
 \label{F6}
\end{figure*}

Computations with different combinations of $\chi$ and $N$ in the
strong segregation range allow us to test the scaling prediction
for the lamellar distance $\lambda = 2\pi /k_\star$ of the form
$\lambda/N^{1/2} \sim (\chi N)^n$.  This relation is obtained by
minimizing the sum of the elastic energy due to molecular
stretching and the interfacial energy within the period $\lambda$,
yielding $n = 1/6$. \cite{Bi94}  Indeed, a good collapse of data
for different $N$ and $\chi$ is achieved in the double logarithmic
plot of $\lambda/N^{1/2}$ versus $\chi N$, as shown in
Fig.~\ref{F6}. The exponent deduced from the slope of the straight
line is $n \simeq 0.22$, in fair agreement with the mean field
argument indicated before. For small $(\chi N)$-values
corresponding to the disordered phase, a saturation is seen, i.e.\
$\lambda \propto N^{1/2}$.

\subsection*{3.2 Diffusion}

\begin{figure*}[ht]
\centering\includegraphics[width=0.90\textwidth]{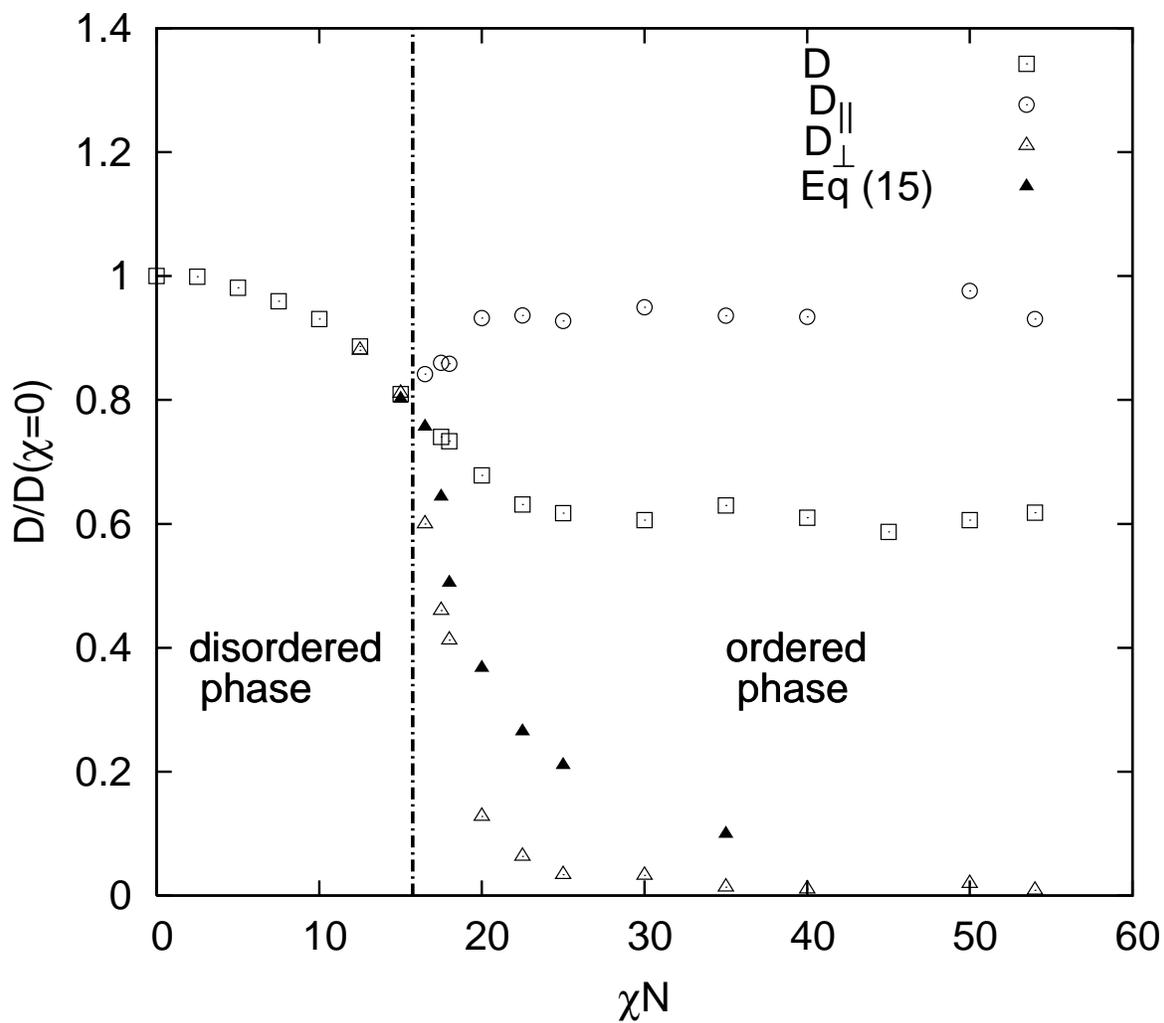}
\caption{Normalized diffusion coefficient $D$ as well as
anisotropic diffusion coefficients $D_\parallel, D_\perp$ in the
lamellar phase, versus $\chi N$. The vertical dashed dotted line
separates isotropic from anisotropic diffusion. Its position
agrees with estimates for the ordering transition based on
equilibrium simulations (section 3.1).}
 \label{F7}
\end{figure*}

 As shown previously,\cite{Mu99,Hof1} the center of mass diffusion
 of molecules in the disordered phase slows down upon increasing $\chi
 N$, because of the increasing degree of short range order, reflected
 by the increase of $S(k_\star)$.  In the lamellar phase, diffusion
 parallel to the interfaces is quite similar as in the disordered
 phase near the ordering transition, whereas perpendicular diffusion
 gets strongly suppressed and drops to zero when $\chi N$ is increased
 further. As seen in Fig.~\ref{F7} the GDM reproduces theses general
 features. To distinguish $D_\parallel$ from $D_\perp$ in the ordered
 phase, oriented lamellae were prepared through appropriate initial
 conditions.  The slight increase of $D_\parallel$ right above the
 transition point may originate from the reduction of structural
 fluctuations and smoothening of the interfaces. The overall diffusion
 coefficient is consistent with the average $D \approx (2
 D_{\parallel} + D_{\perp})/3 \approx (2/3)D_{\parallel}$.
 Analysis of the time dependent center of mass mean squared
 displacements showed that the directionally averaged short-time
 diffusion coefficients are almost unaffected by the interaction
 $\chi$ and are close to $D(\chi = 0)$.

An intuitive approach to understand the diffusion across the
domain boundaries in the lamellar phase is to adopt the picture of
one-dimensional Brownian motion in a periodic potential $V(z)$. We
start from the exact expression \cite{Die77} $D = D_0/ \left[
\overline{e^{-V(z)}} \, \overline{e^{V(z)}}\right]$, where $D_0$
is the diffusion coefficient in the corresponding homogeneous
state, and the bars denote averages over one period. Obviously,
the rate determining diffusion steps are those near the maximum of
$V(z)$ or the minimum of the equilibrium density $\rho_{eq}(z)
\propto \exp (-V(z))$.  Within mean field theories \cite{Die86}
this aspect can be generalized to interacting systems by regarding
$V(z)$ as potential of mean force, defined in terms of the actual
equilibrium density via the Boltzmann factor. Guided by these
ideas, we write for our system
\begin{equation}\label{A1}
D_\perp \simeq \frac{D_0}{\left[ \overline{\rho_A(z)}\,
\overline{\rho^{-1}_A (z)}\right]}
\end{equation}
with simulated $A$-monomer densities $\rho_A (z)$ and diffusion
constant $D_0$ at the ordering transition point. Close to the
ordering transition $\rho_A(z)$ can be represented as $\rho_A(z)
\simeq \overline{\rho_A} (1 + \Delta\rho_A\sin k_\star z)$, which
yields $D_\perp = D_0 (1 - \Delta\rho_A^2)^{1/2}$.  This simple
approach already describes the sharp drop in $D_\perp$ for $\chi
N>(\chi N )_c$, shown in Fig.~\ref{F7}. It still overestimates the
simulated data for $D_\perp$ especially at larger $\chi N$. One
reason may lie in the fact that shape deformations of molecules
during barrier crossing are not included in these arguments. The
barrier crossing problem for a block copolymer has been studied in
Ref. \cite{Barrat}.

 \section*{4. Thin films}

Confinement of block copolymers between planar walls adds several
new aspects to microphase separation.\cite{Bi99,Somm99,kichu94} A
common repulsion of A and B blocks by neutral walls will favor
parallel orientation of molecules and therefore can induce
perpendicular lamellar ordering. However, if walls act differently
on A and B-monomers, one type of blocks will be preferred to the
walls, favoring parallel lamellar ordering. These competing
behaviors, including their time dependence under initial
conditions corresponding to a disordered state at high
temperatures, have been studied before. Hence we limit ourselves
to a few representative and supplementary results.

Patterned walls are particularly interesting for applications,
where a chemically pre-patterned surface should be translated into
a correspondingly patterned polymer
film.\cite{Ro99,Qiang00,Chen98} We show that the GDM can provide
new insight into the process of pattern translation. Calculations
of the time-dependent structure factor enable us to discuss the
propagation of stripe patterns into the film and their competition
with spontaneous ordering fluctuations away from the pre-patterned
substrate.

In the film simulations periodic boundary conditions are used in
the lateral directions.  Unless otherwise specified, films have a
lateral size $L_x=L_y=4 \lambda$ with $\lambda$ the lamellar
distance in the bulk (for example, $\lambda =31$ for $N = 120$ and
$\chi = 0.45$, see section~3). Averages are performed over
typically 10 independent runs.

\subsection*{4.1 Homogeneous walls}

We start with planar walls parallel to the $(x,y)$-plane, with a
potential for A and B monomers as in Ref. \cite{Eu02},
\begin{eqnarray}\label{Vvonz}
V_X^{(1)}&=&\epsilon_X^{(1)}\exp\left(-\frac{z}{2l_w}\right)\,,\nonumber\\
V_X^{(2)}&=&\epsilon_X^{(2)}\exp\left(-\frac{L_z-z}{2l_w}\right)\,,
\end{eqnarray}
where the superscript (1) refers to the lower wall at $z=0$ and
the superscript (2) to the upper wall at $z=L_z$. The quantity
$l_w$ is a parameter that tunes the softness of the
confinement.\cite{Mu99}  For simplicity we have chosen it to be
the same for A and B monomers and the same for both walls,
$l_w=0.5$. The exponential form of the wall-monomer interaction is
convenient, since it allows one to get analytical expressions for
the integrals that determine the interaction with Gaussian monomer
densities.\cite{Eu02}

\begin{figure*}[ht]
\centering\includegraphics[width=0.60\textwidth]{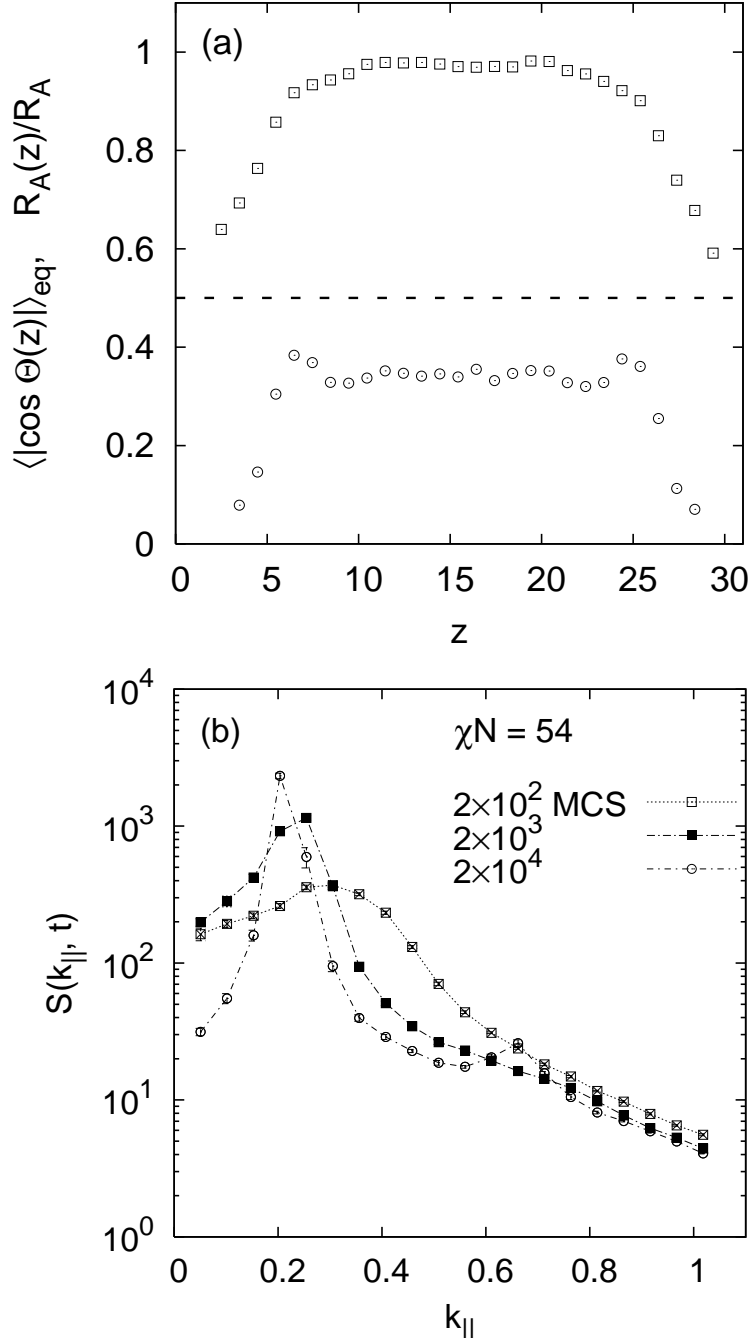}
\caption{(a) Wall-induced molecular orientation (circles) and
normalized radius of gyration $R_A (z) = \langle R_A^2 (z) \rangle
^{1/2}$ (squares) of blocks  across the slab for $\chi = 0.45$.
The horizontal dotted line corresponds to random orientation,
$\langle |\cos \Theta(z)|\rangle = 1/2$. The grid size along the
$z$-axis is $\Delta z = 1$. Chosen parameters are $f_A = 0.5, \chi
= 0.45$  and $N = 120$. (b) Time evolution of circularly averaged
structure factor after averaging over $z$. Note the appearance of
the 3rd-order peak in the final equilibrated state. }
 \label{F8}
\end{figure*}

As discussed above, for neutral walls, molecules near the wall
acquire a preferential parallel orientation. To show this, we set
$\epsilon^{(j)}_A=\epsilon^{(j)}_B=\epsilon_w = 1$, $j=1,2$, and
calculated the quantity $\langle |\cos \Theta(z) |\rangle$ in the
equilibrium state (after thermalization), where $\Theta$ denotes
the angle between the vector $\vec r$ of a molecule and the
$z$-axis.  The results are shown in Fig.~\ref{F8} after averages
were taken over $x$ and $y$ for fixed $z$ coordinate.  Obviously,
angles $\Theta$ near $\pi/2$ prevail near $z=0$ and $z=L_z$. A
preferred parallel orientation persists even up to the middle of
the slab (random orientations would give $\langle |\cos \Theta(z)
|\rangle=0.5$).  The repulsive interactions between monomers and
the walls lead to a deformation of molecules close to the walls,
as seen by the behavior of the average block radius of gyration
versus $z$. At this point the neglect of the internal molecular
structure in the GDM becomes apparent. Close to a wall, chain
polymers have been shown to get compressed in the $z$-direction,
but elongated in the $(x,y)$-plane.\cite{Somm99} Such anisotropic
deformations are beyond the scope of the GDM. This sets a lower
limit of film thicknesses $L_z $ that can reasonably be treated
within the GDM, i.e.\ $L_z$ should be larger than a few times the
radius of gyration.

To further quantify the order in the film and its evolution
 following a quench from a random initial state, we introduce the
 time-dependent lateral structure factor
 \begin{equation}\label{structurefactor}
 S(\vec k_\parallel, z, t) = \frac{L_z}{MN_A}\langle |\rho_A (\vec
 k _\parallel, z, t)|^2 \rangle
 \end{equation}
 defined in terms of lateral density fluctuations
 \begin{equation}\label{lateral}
 \rho_A (\vec k_\parallel , z, t) = \int d^2 x_\parallel \rho_A
 (\vec x ,t) e^{i \vec k_\parallel\cdot \vec x_\parallel}
 \end{equation}
 with $\vec x_\parallel = (x,y)$. The prefactor $L_z$ in
 (\ref{structurefactor}) is introduced in order to achieve
 $L_z$-independence of (\ref{structurefactor}) in the bulk limit $L_z
 \rightarrow \infty$ for given total monomer concentration. Averaging
 (\ref{structurefactor}) over $\vec k_\parallel$ with $k_\parallel =
 (k_x ^2 + k_y ^2)^{1/2}$ fixed and subsequent averaging over $z$
 yields the quantity $S(k_\parallel, t)$ which is plotted in
 Fig.~\ref{F8}b. It clearly reflects lateral ordering induced by the
 alignment effect described above, with some initial
 coarsening. During the coarsening the position of the maximum
 $k_\parallel^\star (t)$ approaches a non-zero value, corresponding to
 the lamellar distance in the equilibrium state. The structure is in
 fact close to the structure found in the bulk in Fig.~\ref{F4}b. In
 particular, a 3rd-order peak appears, reflecting a well ordered
 lamellar structure perpendicular to the walls.

\begin{figure*}[ht]
 \centering\includegraphics[width=0.70\textwidth]{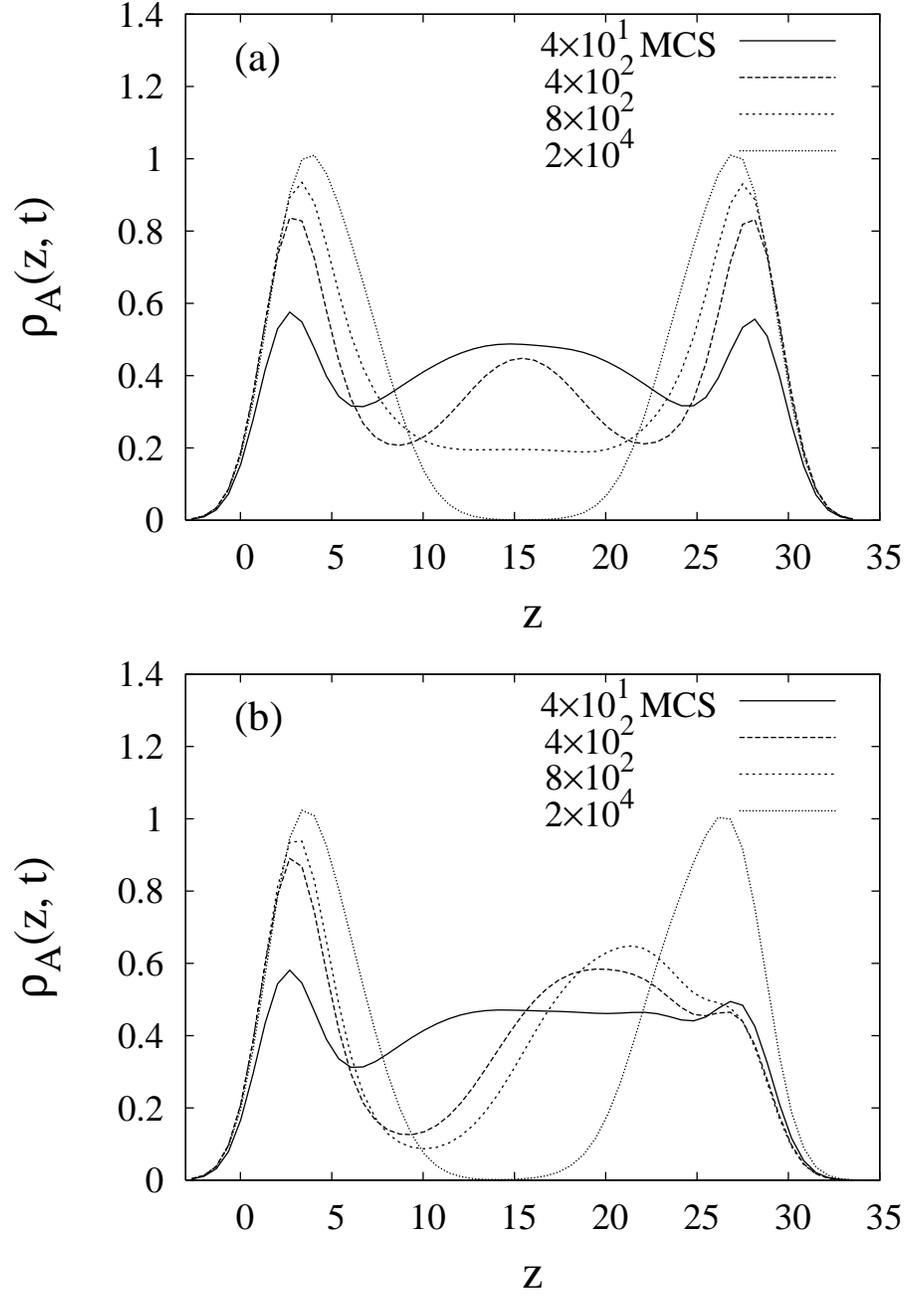}
 \caption{(a) Time evolution of the A-monomer density in a film of
thickness $L_z = \lambda$ with A-attractive walls, for $\chi =
0.45$  (b) Same, but with A-attractive left and neutral right
wall.}
 \label{F9}
\end{figure*}

The situation changes when considering an A-attractive wall,
caused by a stronger repulsion for B than for A monomers,
\begin{equation}\label{Valpha}
\epsilon^{(1)}_X=\epsilon^{(2)}_X=\epsilon_w(1+\delta_w^X)
\end{equation}
where $\epsilon_w=1$ as before and
$\delta_w^B=-\delta_w^A=\delta_w>0$. Choosing $\delta_w=0.5$, the
energetic preference of A-monomers is strong enough to overcome
the essentially entropic alignment effect discussed before that
molecules near the wall have reduced orientational degrees of
freedom. In Fig.~\ref{F9}a the corresponding A-monomer density
$\rho_A(z)$ is plotted across a film of thickness $L_z=\lambda$
(with $\lambda$ from the corresponding bulk simulations). Layers
of A-monomers adjacent to the walls rapidly form and get separated
by a B-rich domain. When the wall at $z = L_z$ is replaced by a
neutral one ($\epsilon^{(2)}_A=\epsilon^{(2)}_B=1$), the
equilibrium density profile essentially remains symmetric, see
Fig.~\ref{F9}b. Thus the A-attractive wall induces layering almost
as in Fig.~\ref{F9}a. Development of the A-rich layer near the
neutral wall, however, takes much longer time than in
Fig.~\ref{F9}a. The whole situation is reminiscent of wall-induced
spinodal decomposition in films of binary polymer blends, at least
in its early stages.\cite{Puri,Fischer,Eudiss}

\subsection*{4.2 Periodically patterned walls}

In this section we study the kinetics of pattern transfer into the
film. The wall near $z=L_z$ is assumed to be neutral
($\epsilon^{(2)}_A=\epsilon^{(2)}_B=1$), while the wall
near $z=0$ has a modulation along the $y$-axis with period $L_p$,
\begin{equation}\label{Vyz}
\epsilon_X^{(1)}(y)= \epsilon_w[1+\delta_w^X\cos(k_p y)]\,.
\end{equation}
Here $\epsilon_w=2$, $\delta_w^B=-\delta_w^A= \delta_w = 0.5$, and
$k_p=2\pi/L_p$.

As long as $\chi < \chi_c$ the equilibrium state displays a
periodic segregation of monomers in the $y$-direction confined
near $z = 0$. The decay of the segregation amplitude along the
$z$-axis reflects the correlation length of the disordered phase.
On the other hand, for $\chi > \chi_c$ various scenarios of
pattern-induced microphase separation emerge, depending on film
thickness and on the commensurability between the two length
scales $L_p$ and $\lambda$.

\begin{figure*}[ht]
\centering\includegraphics[width=0.55\textwidth]{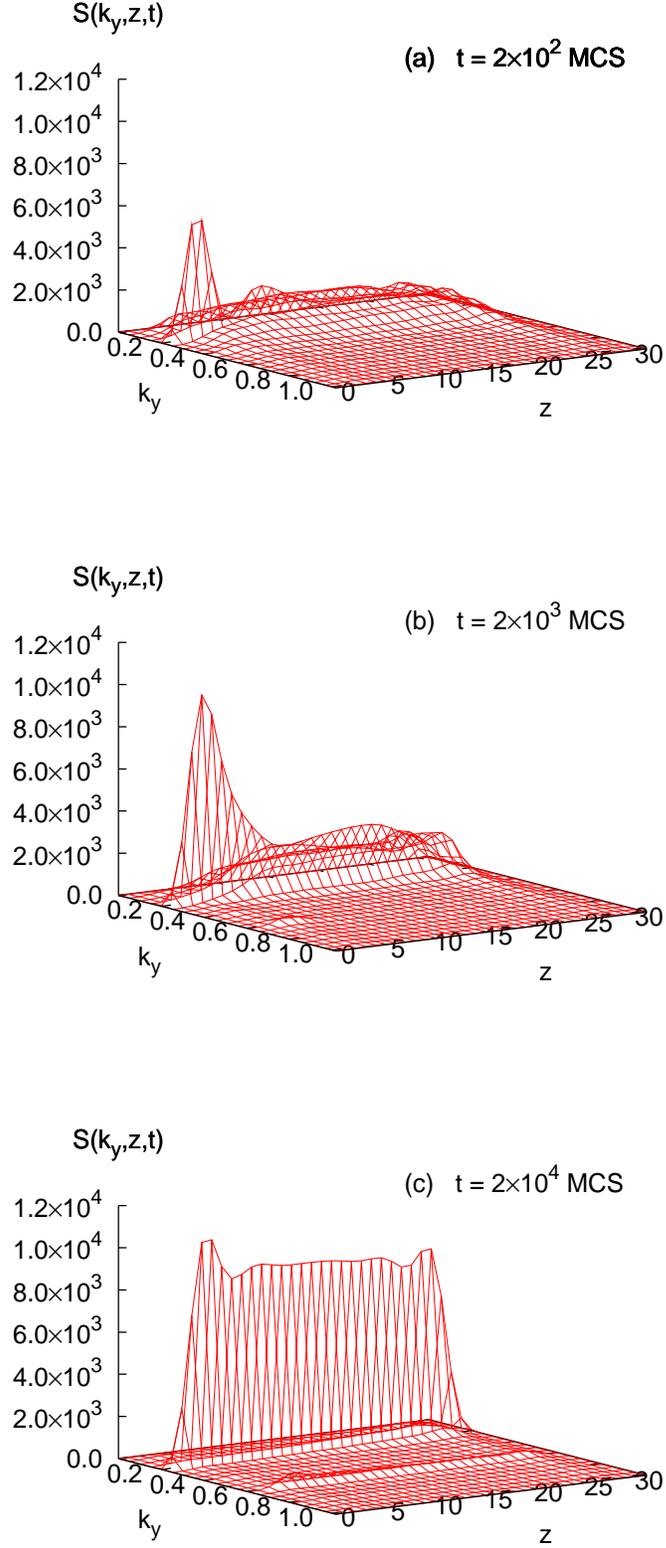}
\caption{Time evolution of the structure factor $S(k_y, z,t)$ in
the presence of a stripe-patterned wall near $z = 0$, for $\chi =
0.45$. The pattern periodicity is $L_p = 2 \pi/k_p = \lambda$ and
the film thickness $L_z = \lambda$. The lateral system size is
$L_x = L_y = 4 L_p$. (a) $t = 200$ MCS, (b) $t = 2\times 10^3 $
MCS, (c) $t = 2 \times 10^4$ MCS.}
 \label{F10}
\end{figure*}

Let us begin with the commensurate case $L_p/\lambda=1$ and a
thickness $L_z=\lambda$. Figure \ref{F10} shows, for $\chi =
0.45$, the time evolution of the integrated structure factor
$S(k_y,z,t)=\int d k_x S (\vec k_\parallel,z,t)$. After a rapid
initial growth of a peak with $k_y \simeq k_p$ near the patterned
substrate, the associated periodic structure propagates along the
$z$-direction. Simultaneously, a broader structure develops across
the slab and indicates spontaneous microphase ordering similar to
bulk behavior.\cite{Fussnote} At $t = 2 \times 10^4$ MCS, the
wall-induced sharp structure dominates everywhere and has
penetrated the slab nearly uniformly. A weak 3rd-order peak ist
clearly seen at $k_y\simeq3k_p$. The result is an equilibrium
state with nearly perfect perpendicular lamellar ordering. No
changes are observed any more when going to larger times.

\begin{figure*}[ht]
\centering\includegraphics[width=0.5\textwidth]{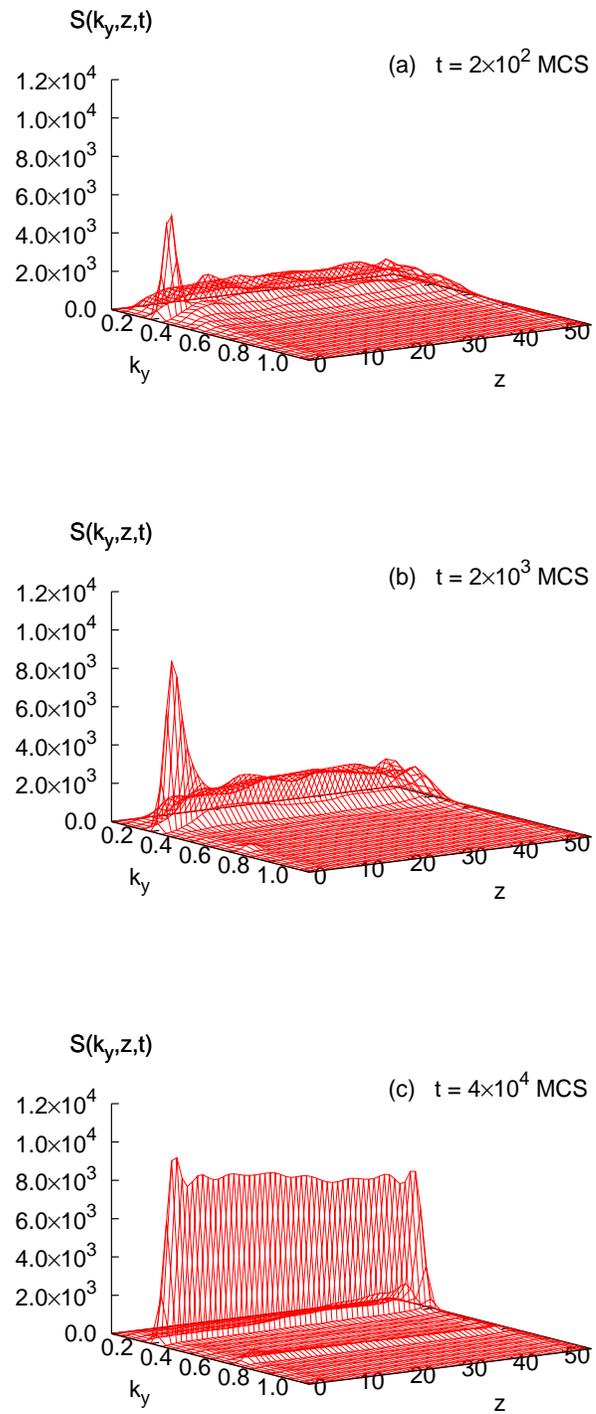}
\caption{Same as Fig. \ref{F10}, but $L_z = 1.8 \lambda$, and $t =
4 \times 10^4$ MCS in (c).}
 \label{F11}
\end{figure*}

For a thicker film with $L_z = 1.8 \lambda$, see Fig.~\ref{F11},
``bulk-like'' ordering processes throughout the slab evolve
further and more strongly interfere with the ordering wave
propagating from the patterned wall. For  long times the ordering
wave finally succeeds to overcome the bulk-like domain structures.
Equilibrium is nearly but not fully reached at $t=4\times10^4$
MCS.

Now we turn to incommensurate situations. The case $L_p/\lambda =
1.1$ and $L_z = \lambda$ will be described  without showing
explicit results because of its similarity at long times to the
commensurate case $L_p/\lambda = 1.0$ shown in  Fig.~\ref{F10}.
The main difference at shorter times is most directly observed
from the structure factor $S(k_{\parallel}, t)$ defined as in
Fig.~\ref{F8}b. In the commensurate case bulk ordering initially
leads to a shoulder in $S(k_{\parallel}, t)$ next to the primary
peak at $k_p$, which merges with it as time proceeds. On the other
hand, for $L_p/\lambda = 1.1$ and $t = 200$ MCS one can identify a
separate ``bulk'' peak. It also merges with the peak at $k_p$ for
longer times $t > 2 \times 10^4$ MCS so that the final structure
corresponds to complete penetration of the wall pattern, as in the
commensurate case. It is interesting to note that transient
coexistence of two peaks and final dominance of the peak at $k_p$
have been observed recently in two-dimensional Fourier transforms
of real space images of diblock copolymer thin films on chemically
nanopatterned substrates.\cite{Edwards}

\begin{figure*}[ht]
\centering\includegraphics[width=0.5\textwidth]{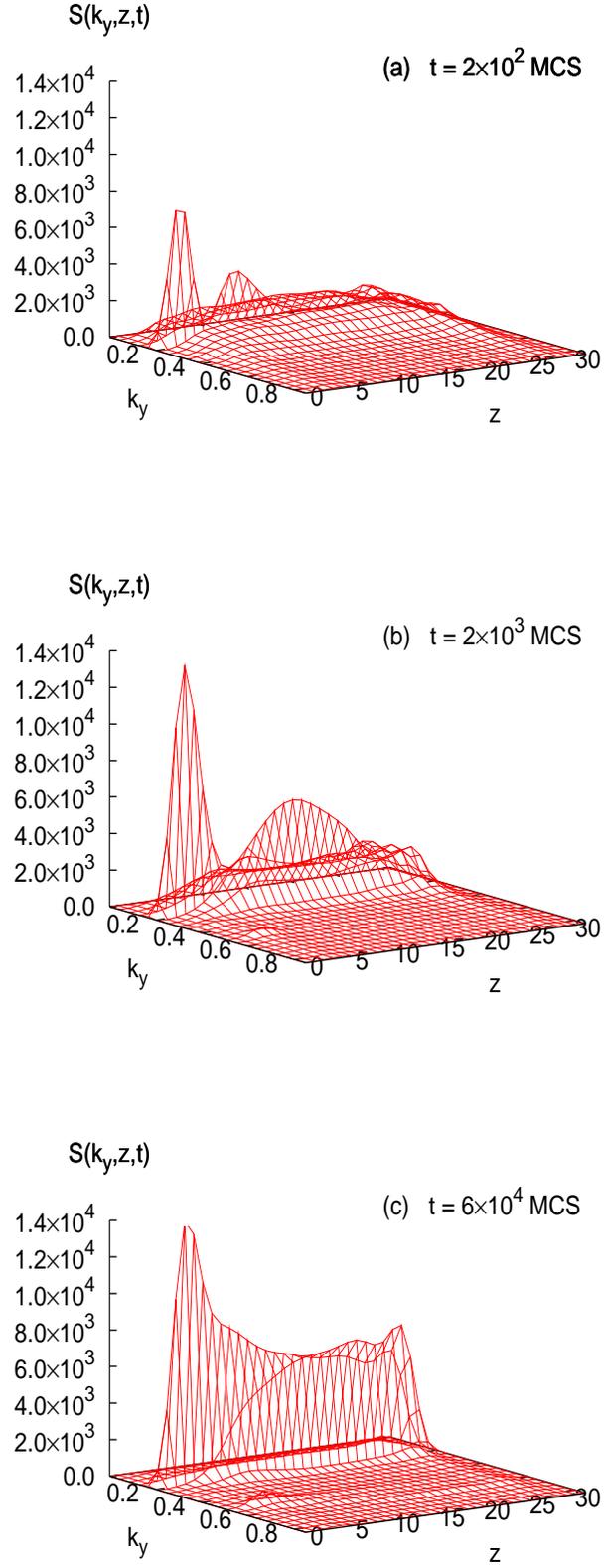}
\caption{Same as Fig.~\ref{F10}, but $L_p = 1.2\lambda$, and $t =
6 \times 10^4$ MCS in (c).  }
 \label{F12}
\end{figure*}

\begin{figure*}[ht]
\centering\includegraphics[width=0.90\textwidth]{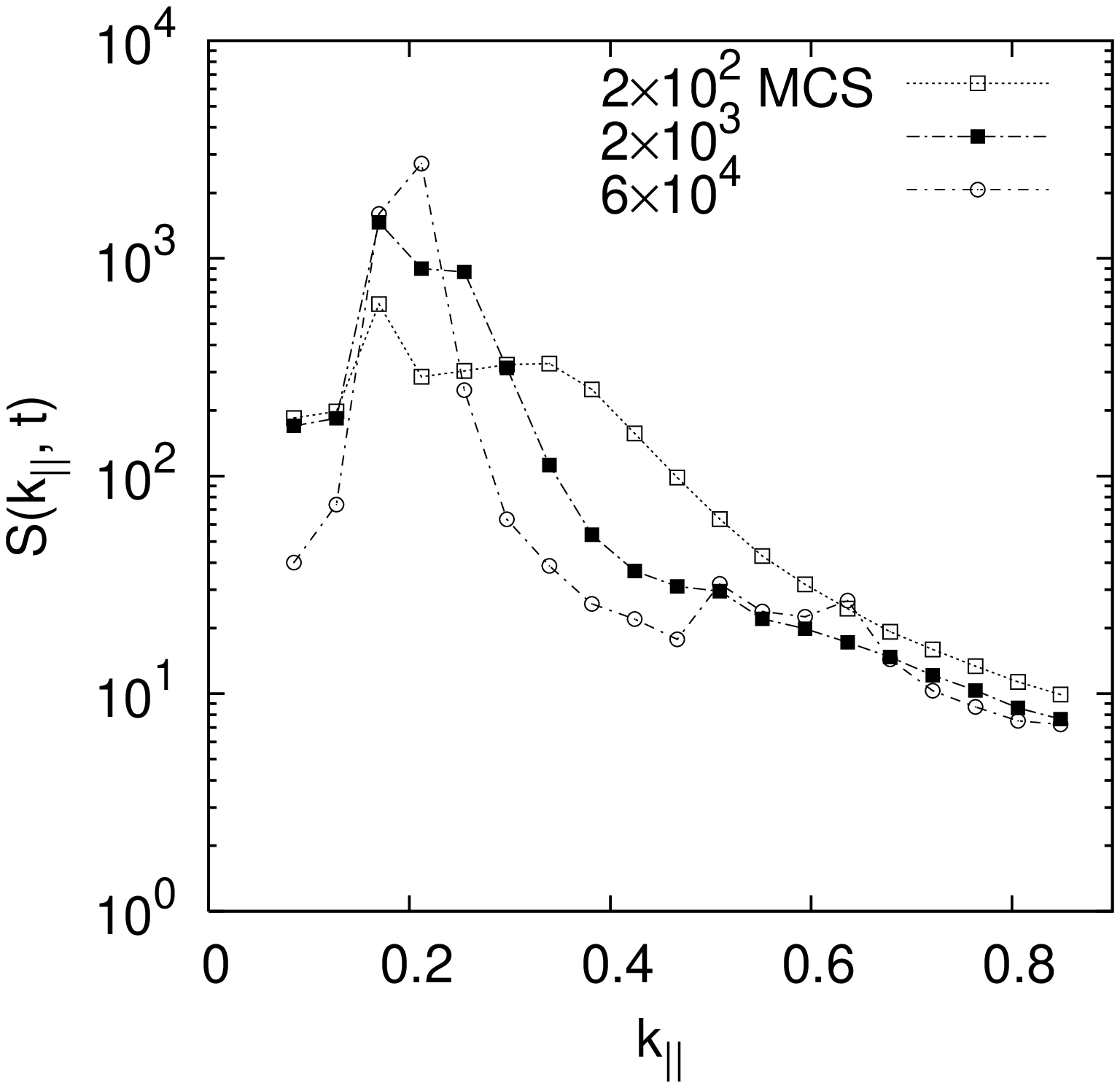}
\caption{Circularly and $z$-averaged structure factor from the
same simulation data as in  Fig.~\ref{F12}.}
 \label{F13}
\end{figure*}

By contrast, for the larger ratio  $L_p/\lambda = 1.2$ with $L_z =
\lambda$ the two main features in Figs. \ref{F12}a - c prevailing
near the patterned wall and deeper in the film, respectively,
remain separated with respect to $k_y$ as time proceeds. Up to $t
= 2 \times 10^3$ MCS the peak at $k_p$ dominates, which is evident
from the averaged structure factor $S(k_{\parallel},t)$ plotted in
Fig.~\ref{F13}, whereas for $t > 2 \times 10^4$ MCS the quantity
$S(k_{\parallel}, t)$ takes its maximum at the bulk value $k_0 = 2
\pi/\lambda$. No further change of this pattern is observed in our
longest runs up to $t = 6 \times 10^4$ MCS, nor does it change
when we increase the strength in the modulation of the wall
potential to $\delta_w = 0.75$. Hence it appears that no
well-ordered structure develops in this case, showing that
transfer of the wall pattern to the film sensitively depends on
the commensurability of the two length scales $L_p$ and $\lambda$.

\section*{5 Summary and Outlook}

The Gaussian disphere model (GDM) provides a highly coarse grained
description of diblock copolymer melts at the molecular level
which still captures the essential features of self-organized
structure formation. The essence of the model is to parameterize
the internal molecular degrees of freedom by a few stochastic
variables, the radii of gyration $R_A$ and $R_B$ of each block and
the distance vector $\vec r$. Molecular positions together with
these internal variables move according to a kinetic MC algorithm.
The algorithm is driven by a free energy $F (\{\vec
r_i\},\{R^A_i\},\{R^B_i\})$ derived from Gaussian chains, in
analogy to the Gaussian ellipsoid model for polymer melts proposed
earlier.\cite{Eu01,Eu02}

Regarding bulk ordering and diffusion, several features known from
less coarse grained models were reproduced to a good
approximation. This prompted us to study microphase ordering in
thin films, especially the kinetics of pattern transfer from a
stripe-patterned substrate into the film. Detailed results for the
time-dependent structure factor $S(k_y , z, t)$ were presented,
including some discussion of commensurability effects and pattern
penetration into films of varying thickness.

The GDM is expected to loose reliability, when the film thickness
becomes less than a few times the radius of gyration. Then the
assumed sphericity of individual blocks is no longer compatible
with the molecules' distortion under confinement. Improvement on
that issue could be achieved in the spirit of Ref. \cite{Eu02} by
allowing ellipsoidal block shapes or by representing each block as
a string of soft spheres.

On the other hand, the GDM may turn out advantageous in a
semi-quantitative description of more complex molecules, including
chain-like segmented or branched copolymers, or copolymers
carrying reactive groups.\cite{Krakovsky}

\section*{Acknowledgement}Financial support by the Deutsche
Forschungsgemeinschaft (International Research Training Group
``Soft condensed matter'') and the UFA (Universit\'{e} Franco
Allemande) is gratefully acknowledged.

\newpage
\begin{table}[hbtp]
\centering \caption{Parameters for fitting the generalized Leibler
function (\ref{NS1}) to  structure factor data in Fig. \ref{F2}.}
\begin{tabular}{||c|l|l|l|l|l|l||} \hline \hline
  $\chi N$        & 0.0  & 2.5  &  5.0  &  10.0  & 12.5  & 15.0\\ \hline \hline
  $\alpha$        & 1.9 & 1.8 & 1.9 &  1.6  & 1.4  & 1.5\\ \hline
  $\delta$        & 2.1 & 6.4 & 8.9 &  15.5 & 17.9 & 19.2\\ \hline
  $\tilde R_G$  & 6.1 & 6.1 & 6.3 & 6.3  & 6.4  & 6.5\\ \hline
\end{tabular}
\end{table}

\newpage
\newpage

\newpage

\newpage

\newpage

\newpage

\newpage

\newpage

\newpage

\newpage

\newpage


\newpage

\newpage

\newpage

\begin{thebibliography}{99}
\bibitem{Fussnote} At early times a smaller peak in $S(k_p, z,
  t)$ and a well defined minimum near $z \approx 2 \langle (R^A)^2\rangle^{1/2}\approx 7.2$ can
  be observed in between these two main features. This secondary
  peak results from AABB block sequences (at fixed $z$) due to
  molecules in the second layer from the wall, with preferential
  perpendicular orientation. This is inferred from the
  $z$-dependence of $\langle |\cos \Theta (z)|\rangle$.
\end{thebibliography}
\end{document}